# SMART STAMP DUTY

Dimaz Ankaa Wijaya[a], Fengkie Junis[b], Dony Ariadi Suwarsono[c]

[a] Monash University, Australia Email: dimaz.wijaya@monash.edu
[b] Universitas Gadjah Mada, Indonesia Email: fengkie.junis@mail.ugm.ac.id
[c] Direktorat Jenderal Pajak, Indonesia Email: dony.suwarsono@pajak.go.id

**ABSTRACT**

*Blockchain technology has enjoyed a massive adoption in cryptocurrencies such as Bitcoin. Following the success, many people have started to explore the possibility of implementing blockchain technology in different fields. We propose smart stamp duty, a system which can revolutionize the way stamp duty is managed and paid. The smart stamp duty offers significant improvements on the convenience when paying stamp duty. At the same time, the blockchain technology also provides the auditability of the transaction data. Smart stamp duty enables the expansion of the existing electronic stamp duty application to retail level as well as allows the taxpayers to pay the stamp duty of their electronic documents. Our proposed system also enables the taxpayers to print their electronic documents without losing the paid electronic-based stamps.*

*Teknologi blockchain telah menikmati sukses besar di bidang mata uang kripto seperti Bitcoin. Sukses ini memacu banyak orang untuk mengeksplorasi kemungkinan implementasi teknologi blockchain dalam bidang lain. Kami mengajukan ide* smart stamp duty *(bea meterai pintar), sebuah system yang dapat merevolusi cara bea meterai dikelola dan dibayar. Bea meterai pintar menawarkan peningkatan signifikan terhadap kemudahan membayar bea meterai. Pada saat yang bersamaan, teknologi blokchain juga menawarkan auditabilitas informasi transaksi. Bea meterai pintar memungkinkan perluasan aplikasi bea meterai elektronik ke tingkat pengguna individu yang sekaligus memungkinkan para pembayar pajak untuk menunaikan kewajiban bea meterai terhadap dokumen elektronik mereka. Sistem kami juga memungkinkan para pembayar pajak untuk mencetak dokumen elektronik mereka tanpa kehilangan bukti bayar bea meterai elektronik.*

KEYWORDS: stamp duty, blockchain, smart contract, Indonesia, electronic documents.

## 1.    INTRODUCTION

According to 2018 Indonesia State Budget, taxation is the major source of state revenue, comprising 1.6 quadrillion Rupiah. The number has sharply increased from 1.47 quadrillion Rupiah in 2017 [1, 2, 3, 4] . The increase of taxation target in 2018 is due to several assumptions, including automatic exchange of information (AEoI), tax incentives, and improvements around human resources, information technology, and taxation services to the taxpayers.

Stamp duty is one of the taxes in "Other Taxes" category implemented in Indonesia, which is targeted at 9 trillion Rupiahs or about 0.6% of the overall 2018 revenue, increasing 11% from the 2017 state budget. Despite the small percentage the stamp duty carries, Directorate General of Taxes (DGT) as the Indonesian Tax Authority will also look for revenue improvements in this area   [5]. Online transactions should be evaluated as the users might also need to pay stamp duties on the electronic documents, aside from the paper-based documents, involved in the business [6].

### 1.1.    Problem Definition

Currently, stamp duty's taxable object only focuses on paper-based documents, which is clearly defined on the Stamp Duty Act 1985 [7]. The taxpayers need to attach physical stamps, sign them properly, write the dates, and only then the tax payments are deemed valid. The problem with physical stamp is that it is inconvenient to get, as the taxpayers need to buy the physical stamps on any brick and mortar stores to be used on their printed documents.

There is also a limited use of computerized stamp duty which is only valid under a certain circumstance [8]. To use the computerized stamp duty, the taxpayers need to pay the tax due up front according to their estimation of the number of documents they will create in the following month. It is also a requirement that the minimum number of stamped documents is 100 per day.

The existing stamp duty mainly focuses on physical documents; hence the physical stamp is required. However, as technology advances, computerization changes the way people conduct their business. Electronic documents are more convenient to create, manage, and process compared to physical documents. Although these electronic documents may be owed stamp duty, it is infeasible for the taxpayers to purchase physical stamps and attach them to the electronic documents [9].

In summary, we identify the gap of using computerized stamp duty where individuals or companies that prefer to use electronic documents where these documents are taxable by the stamp duty regulations. Several solutions have been proposed to solve the problem [10, 6]. However, the usability and correctness of the data stored in the proposed solutions are questionable, where the payment proofs are stored in an information system without any additional features.

### 1.2. Our Contribution

Considering the current limitation of the stamp duty, we propose a new blockchain-based electronic stamp duty as an effort to increase the stamp duty revenue by technological advancement. The system will increase the convenience of the taxpayers to pay the stamp duty when creating taxable documents. The system can be used by all taxpayers by using information technology devices. The system will also provide a public verifiability for proving that a certain electronic document has the stamp duty paid. The built-in cryptographic techniques also prevent forgery which happened in the physical stamp duty [11, 12].

We coin the term **smart stamp duty** to refer our proposed system which utilises smart contract as well as blockchain technology to record information and conduct predetermined business processes automatically based on the user's inputs and program logic. The smart contract is fully auditable, transparent to its stakeholders, and the finality of the data is guaranteed. Moreover, its standard features such as timestamp and digital signature fulfil the requirements of a valid stamp duty. Smart contract is scalable; not only retail users but also wholesale users can use the smart stamp duty as long as there exist sufficient applications built to communicate with the smart contract.

## 2. BACKGROUND

In this chapter, we discuss several background topics related to our proposed solution, namely taxation, stamp duty, and existing blockchain technology solutions in taxation area.

### 2.1. Electronic Stamp Duty

Stamp Duty Act 1985 determines that there are documents (in the physical form) owed stamp duty [7]. As a proof that the tax is paid, a form of physical stamp is glued on the documents and signed. The cost of the stamp depends on the category of the document and the amount of money printed on the document.

The business model of electronic stamp duty is investigated by [10, 6]. The Information and Electronic Transaction Act 2008 has provided a foundation in which the term document can be extended, not only in the physical form but also in electronic form, since both forms are now considered as legal proofs [13]. Before the act was established, all legal proofs must be printed and legalized, but now all electronic information no longer need to be printed in physical form, and if the information is properly handed, it can be a legal proof.

## 2.2. Blockchain

Blockchain was first applied in Bitcoin, a decentralized payment system introduced by Satoshi Nakamoto [14]. The blockchain technology enables users to create transactions without the need of proving their identities. There is no central authority controlling the system, instead every user can verify and validate transactions data by using a set of protocols.

Blockchain can be divided into three models based on its type of participants, namely public blockchain, private blockchain, and consortium blockchain [15]. Public blockchain is a type of blockchain which allows everyone to participate at any time and allows anyone to leave whenever they desire. In a private blockchain, as its name implies, is a closed system where the participants need specific permissions from a central authority which controls the whole system. Although blockchain is originally intended to remove the central authority, however in some cases the central authority is still required and cannot be removed, for example due to legal requirements.

The last blockchain model is consortium blockchain. It is a combination between public and private blockchain, where there will be several authorities sharing the power equally among them to control the system, however the membership of the authorities is not open as in a public blockchain. Consortium blockchain also applies access control mechanism where not everyone has access to the blockchain.

To replace the central authority, blockchain utilises consensus mechanism such that the permission to write new information to the blockchain is determined by a consensus method which is run by the participants. Consensus mechanism is applicable to public blockchain and consortium blockchain, while it is uncommon for the private blockchain to have any consensus mechanism.

## 2.3. Smart Contract

While blockchain as in Bitcoin has a limited set of operation codes (opcodes), smart contract platforms such as Ethereum was created to support Turing-complete application in the blockchain [16]. Smart contract is generally an application stored in the blockchain. It receives inputs from the user, runs functions, and generates outputs (if any). As it is stored in the blockchain, the codes are permanent and transparent, hence it offers fairness to the participants. A smart contract platform also inherits the same traits as other blockchain systems, where the transaction data is visible and verifiable to anyone.

## 2.4. Cryptographic Techniques

Blockchain technology relies on cryptography to run the system. Some of the techniques are public key cryptography (PKC) and hash function (HF). The PKC enables the two different keys, namely public key and private key. The public key will be the address where a user stores and sends tokens or coins, while the private key will be used to create digital signature as a proof of ownership of the digital asset. HF computes an arbitrary length data into a fixed-length value as its output, called Hash Value (HV). HV is commonly used to simplify the integrity check of the data.

## 2.5. Blockchain Technology in Taxation

Blockchain is not only showing its potential in the payment industry, but also other areas such as taxation. Previous investigations describe that blockchain technology can mitigate the problem of tax losses due to international trades [17, 18, 19, 20]. The blockchain technology has been proposed to modernize Value-Added Tax (VAT) [21]. In this case, the blockchain is utilized to enhance the information openness across multiple authorities and likewise, to share information between those authorities.

## 3. SMART STAMP DUTY

This chapter describes smart stamp duty, our proposed solution for automating stamp duty mechanisms through smart contract-powered blockchain technology. We identify related participants, blockchain requirements, business processes, as well as the program logics that need to be implemented to the smart contract.

### 3.1. Overview

Our system works in the blockchain environment where cryptographic techniques such as digital signature and hash functions exist. We will define the term tokens to refer to electronic money balance owned by a user. The tokens can be used to pay the stamp duty in a smart contract provided by the Tax Authority. Before a user pays the stamp duty, she needs to buy the tokens before paying the stamp duty for each taxable document she creates. The transparency of the system enables the actors of the system to verify and validate all information stored in the blockchain.

### 3.2. The Blockchain

The blockchain as a shared database which can be used to store the data as well as the smart contracts, and then run the smart contracts based on the input provided by the users. In the proposed system, we use the consortium-based permissioned blockchain model where the Tax Authority along with other trusted participants such as government's auditing bodies can validate new transactions and secure the blockchain.

There will be multiple nodes running the system. These nodes can be managed by multiple participants, for example banks, the tax authority, and the monitoring bodies. Each of these participants keeps a copy of all transactions in the system for auditability and accountability purposes. Although there can be many nodes owned by different participants, the access rights will still need to be set up by the Tax Authority as the system administrator.

### 3.3. The Smart Contract

The smart contract is a set of permanent codes stored in the blockchain. The smart contract can be used to store information and run application logic based on the codes. The purpose of the smart contract is to make the application logic transparent, auditable, and all the information is visible to all users. The smart contract will have several parameters to be set by the owner based on the regulations. The smart contract will be created and owned by the Tax Authority.

### 3.4. The Actors

We determine several actors involved in the system: the users, the bank, the Tax Authority, and monitoring organisations. The actors have their own activities and credentials to do different things.

### 3.4.1. The Taxpayers

The taxpayers are using the system to pay their tax dues, e.g. stamp duty. They interact with the system by using application interfaces provided. The taxpayers interact with the banks or any retail sellers to buy balance by using electronic money, credit cards, bank accounts, or cash. The users produce the documents, pay the stamp duty, and send the documents to other users. The users can also verify that the documents they receive have the tax paid.

### 3.4.2. The Bank

The bank usually provides financial services to the users. In the proposed system, the bank also provides a gateway to convert the local currency to tokens to access the system. The bank ensures that the exchange system is accessible by the taxpayers whenever they want to buy the tokens to be used to pay the stamp duty. At the end of a predetermined period, the bank reports to the tax authority regarding the amount of money collected from the taxpayers and send all the money to the government's account.

### 3.4.3. The Tax Authority

The Tax Authority has the authority to determine the tax rate to be paid by the taxpayers. In our system, the Tax Authority controls the blockchain system, including deploying or modifying the smart contracts used to run the system. The Tax Authority could authorize other parties to get involved in the system.

### 3.4.4. Tax Monitoring Organisations

The tax monitoring organisations help running the system by participating the closed consensus mechanism. In this mechanism, these organisations along with the Tax Authority determine the transactions to be stored in the blockchain. These organisations can have access to the system and the information stored in it.

### 3.5. The Applications

The applications are the interface between the users and the blockchain/smart contract. The applications used by different actors can be different based on their own authority. The application used by The Tax Authority may have the ability to modify the variables (by using the appropriate private keys), but the users can only add new records.

There can also be some additional applications to conduct the business, such as word extractor which enables the user to create the appropriate Bloom Filter to supply the stamp duty payment. Bloom Filter checker will also be required for validation in case the user prints the electronic documents and the printed documents require verification. However, these additional applications are not discussed in this paper.

### 3.6. The Token

Tokens are numbers stored in a smart contract and can be moved from an address to other addresses. The tokens are not intended to replace the fiat currency (Rupiah) as it will violate Currency Act 2011. Tokens are used as proofs that the taxpayers have paid the stamp duty in the form of balance of addresses and transactions. Only one valid token will be created and available in the smart stamp duty system.

### 3.7. Core Business Process

The core business process can be described as in the Fig. 1.

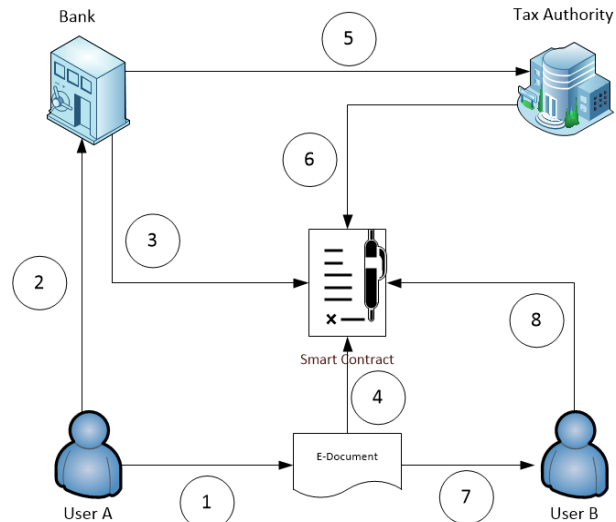
*Fig. 1. Smart Stamp Duty Business Process*

Below is a simple description for each activity in Fig. 1.
1) A user (User A) creates an electronic document (e-document) owed stamp duty. The electronic document can be in any form (word processor, spreadsheet, raw data, etc).
2) User A purchases tokens from a bank which can be used to pay the stamp duty. The user informs the bank of her "address".
3) The bank credits the tokens to the user's address by sending the correct transaction to the smart contract. The token balance can be checked by User A.
4) User A makes sure that the e-document format is unchangeable, then computes the hash value of the e-document, creates a transaction to put the hash value of the document to the smart contract. The transaction will decrease User A's token balance based on the stamp duty paid.
The original document (which has been signed) needs to be kept by User A, while a new copy of the original document can be modified to include the receipt of the transaction.
5) The bank routinely provides a report to the Tax Authority regarding the total amount of balance purchased by taxpayers. The report can also be sent automatically or by using a specific interface.
6) The Tax Authority rechecks the report received from the bank to the actual transactions in the smart contract. The process can also be converted into an automatic job.
7) User A sends both the original copy and the paid copy of the electronic document to User B.
8) User B checks by querying the blockchain that the stamp duty has been paid.

New users can join the system immediately by creating new public key pairs. All transactions can be read by all actors, thus reduce the fraud risks.

### 3.8. The Parameters

Based on the business process we have determined, we could then provide a detailed data to be stored on the smart contract. There are 3 types of the data: system parameters, account parameters, and transaction parameters.

### 3.8.1. Account Parameters

Account parameters are information regarding the token balance of the users. A new account will be created for a new user containing at least the user's public key and an initial token balance (can be zero). The public key cannot be modified while the token balance will always be recalculated by the smart contract based on new transaction parameters.

### 3.8.2. System Parameters

System parameters are information required to run the system. The smart contract looks up to the system parameters before executing transactions. The smart parameters are determined by the Tax Authority based on the existing regulations. The parameters can always be changed whenever the regulation changes.

The system parameters contain at least the following:
- The cost of the stamp.
- The regulation references.

The system parameters can be stored on the same script of the main smart contract or be put on a different script.

### 3.8.3. Transaction Parameters

Transaction parameters are information regarding stamp duty transaction created by the users. To satisfy the existing regulation on stamp duty, the information items required for a valid transaction are the following:
- The public key of the taxpayer.
- The amount of the tax paid.
- Timestamp of the document.
- Hash value of the document.
- The payer's signature.

In the real world, it might be possible for the taxpayers to create identical documents multiple times, hence they produce an identical hash value. Thus, to minimize the occurrence, it is possible to add unique information into the documents, such as public keys of the taxpayers or their tax file number and the document creation time.

### 3.9. The Program Logic

Based on the business process and the parameters, it is possible to determine the program logic needed to run the system. All the program logics are run by the smart contract independently.

### 3.9.1. System Parameter Definition

If the regulation changes, then the Tax Authority can change the system parameter by sending a new transaction. The smart contract checks if all required data is satisfied by the new transaction then stores the new parameter. The system parameters are defined in section 3.8.2.

### 3.9.2. Token Transaction

Activities related to tokens are: token creation, token distribution and token sales, and lastly stamp duty payment.

**Token creation**. As a central controller, Tax Authority is the only party that is given the authority to create tokens. The number of tokens to be created depends on the need. It is also possible to "mint" or create more tokens as required. However, during contract creation, token amount can also be initialized.

**Token distribution and token sales**. Based on existing regulations, the Tax Authority can distribute the tokens to authorized participants such as banks and post office. The authorized participants can sell the tokens to any secondary markets such as distributors and retailers or sell them directly to end users (taxpayers).

**Stamp duty payment**. If the user wants to pay the stamp duty, then the user creates a new transaction to do so. The smart contract first checks if there is enough balance in the user's account, checks if required data is supplied. If everything is satisfied, then the smart contract reduces the amount of the stamp duty to be paid in the user's balance before storing the document data in the smart contract. To prove that the user has paid the cost, the tokens will be sent to the Tax Authority's address. The tokens will be accumulated by the Tax Authority which will be used for audit or reconciliation when calculating the real number of tax paid in fiat currency.

### 3.9.3. Tax Payment Audit

For the purpose of audit, the tokens received by the Tax Authority will not be recirculated. When the audit is finished, then the same number of tokens verified to be the stamp duty revenue could be sent to an unusable address (an address without known private key) such as zero address.

The scope of the audit can be expanded such that auditing bodies can directly assess and monitor the system and evaluate the data in a real time. By using blockchain as a shared ledger, the amount of token that has been sold and the amount of token that has been spent will be easy to calculate.

## 3.10. Transaction Fee

Assuming the blockchain will run in a permissioned environment, then it is possible to waive transaction fees for all transactions created within the system. It is assumed that the Tax Authority will not abuse the system by flooding the system by sending system parameter-defining transactions. It is also assumed that when creating transactions, the Bank and the User need to have enough balance to pay the stamp duty. Without a proper information supply, the transaction will not be processed and confirmed by the system.

However, without any transaction fee, the system is prone to abuse by participants, for example creating many transactions to their own addresses to slow down the entire system. To avoid the useless transaction flooding, a small transaction fee can be introduced such that creating many transactions will be costly to the attacker. The implementation of transaction fee is beyond the scope of this paper.

## 3.11. Security Model

We define the security model of our proposed system as follows. A malicious user wants to create her own tokens without buying from the bank. A malicious bank tries to reduce the report provided to the Tax Authority to keep the tax money illegally.

In the system where the Tax Authority is assumed to be always honest and no security vulnerability is found on the blockchain system and assuming that the cryptographic techniques are secure, then the probability of the malicious user trying to create new tokens is negligible, while the probability of the malicious bank modifying the report without being detected is also negligible.

## 4. IMPLEMENTATION AND EVALUATION

We have implemented our solution which can be deployed on Ethereum blockchain. Our solution consists of two parts: the smart contract and the user interface. The smart contract code is available on Github[1]. The user interface was written in Python and it is also available on Github[2].

The smart contract uses a standard ERC-20 as its foundation which adds the capability of being traded on ERC-20 token markets due to its compatibility. We call the token `Smart Stamp Duty` (`SSD`). As with other ERC-20 tokens, our `SSD` is transferrable. Before paying the stamp duty, the user needs to buy the `SSD` token. Aside from `SSD` token, the native token Ether also exists in Ethereum environment. The transaction fee is paid by using Ether, while the amount of the fee is determined by calculating the complexity of the smart contract, expressed by gas. The gas is then converted into Ether by determining a conversion rate called gas cost or gas price. For simplification, the gas cost can be set to 0 Ether such that paying Ether will not be necessary in the implementation.

The main component, class `StampDuty`, contains two important data structures to contain information related to our solution, namely `StampParam` and `PayParam`. Each structure will be further explained.

### 4.1. Implementing Data Structures
### 4.1.1. StampParam

`StampParam`, contains information about types of stamp duty. Currently, there are only two stamps available which have different denominations: "Rp3000" and "Rp6000". Below is the detail of the `StampParam` structure.

- `StampCode`, to store stamp primary key (unique).
- `StampName`, to store human-readable stamp information.
- `StampPrice`, to store the price of the stamp.
- `RegulationReference`, to store the regulation reference of the stamp.
- `IsActive`, to flag active and inactive stamps.

StampParam is the master table for the stamp duty. Its importance is shown by only allowing the owner (i.e. the Tax Authority) to add and modify the data.

### 4.1.2. PayParam

The second structure, `PayParam`, contains stamp duty payment data. A user adds a `PayParam` information every stamp duty paid. The PayParam structure is shown below.

- `PayCode`, works as the primary key of the payment (unique).
- `DocHash`, stores the hash value of the document related to the paid stamp.
- `PayIndex`, to store the index of the payment.
- `Payer`, to store the payer address.
- `StampCode`, to refer to the stamp paid by the transaction.
- `BloomFilter`, to store Bloom Filter information for testing.
- `TimeStamp`, to store timestamp in integer format.
- `TimeStampStr`, to store timestamp information in string format.
- `PayerSignature`, to store the payer's signature.

As a unique key, the `PayCode` can be generated by a hash function `H` of multiple information, such as `DocHash`, `Payer`, `StampCode`, and `timestamp` when the document is submitted such that:

---

[1] https://github.com/sonicskye/smart-stamp-duty
[2] https://github.com/sonicskye/smart-stamp-duty-ui

$$\texttt{PayCode = H(DocHash||Payer||StampCode||Timestamp)}$$

where the symbol (`||`) is concatenation operation.

`PayerSignature` is generated by a signing function `SIGN` using Ethereum's library under the following formula:

$$\texttt{PayerSignature = SIGN (DocHash, PrivateKey)}$$

Where `PrivateKey` is the private key owned by the payer, associated to the payer's address which is stored in `Payer` variable.

Aside from the two data structures, there are also built-in functions to manage the system, including token minting and token burning to control the token supply.

`BloomFilter` [22] value of the document is also recorded on the smart contract. It supports content matching mechanism when printing the electronic document into paper-based document. The value `BloomFilter` can be used to verify that the electronic and paper-based versions are identical. When computing the `BloomFilter`, the system will list all words in the document under a parameter set, including number of words. The verification is done by sampling keywords from the document. If the verification passes, then we can conclude that the paper-based document is identical to the electronic version of it.

### 4.2. Implementing Content Matching

The content matching feature was implemented in our source code. The feature simply tests a set of input strings which comes from a document against a Bloom Filter value taken from the claimed electronic stamp duty. The input can either be done manually or automatically, where the topic is beyond the scope of the paper. The content matching protocol is as follows.

1) Determine the pair document `W` and the Bloom Filter value `BF` which can be retrieved by querying the smart contract using `PayCode`.
2) Extract all distinct words `w` from the input document `W` such that `W` = {$w_1$, $w_2$, $w_3$, …, $w_n$} where `n` is the number of distinct words.
3) For each word `w`, check whether `w` ∈ `BF`. Store the result in `R`.
4) The result is the percentage of `True` in `R` against the value `n`.
5) The expected value is 100% which determines that there is a possibility that the document `W` is identical to the original document.

### 4.3. Evaluation

We deployed our implementation by using Truffle Framework[3] and Ganache[4] as the development environment in a private Ethereum network. The `PayParam` contains data duplicates such as timestamp and the payer's signature which are actually embedded in the transaction data and in the block where the transactions are included. However, in order to read the timestamp and signature directly from the block requires enormous computing power such that the Ganache which we used as a node in our development environment crashed every time the code was tested. That is why `PayParam` explicitly includes the additional information. As the result of this extra information, the required transaction fee doubled from around `235,000` gas to more than `450,000` gas.

---

[3] https://truffleframework.com
[4] https://truffleframework.com/ganache

## 5. CONCLUSION

We have described a new blockchain-based electronic stamp duty in the system we call **Smart Stamp Duty (SSD)**. The proposed system is expected to expand the tax base for the stamp duty as it enables electronic documents to be stamped. The benefit of our solution is threefold: increasing convenience for the taxpayers to pay the stamp duty, increasing state revenue in the stamp duty sector, and increasing the transparency and auditability of the stamp duty mechanisms. Our solution also enables the format conversion of the document from electronic format into paper-based format where the integrity of both documents can be verified.

## 6. FUTURE WORKS

We plan to evaluate the privacy issue of our proposed smart contract system. It is possible that the users' privacy is analysed based on the pattern of the transactions, thus expose risks on their privacy. We will develop mitigation strategies based on our analysis and implement them on the next version of our design.

We also plan to evaluate the consensus model to avoid a single point of failure that the security of the blockchain relies on the honesty of the Tax Authority. We can introduce new actors to the system or distribute the authority to other actors. Byzantine Fault Tolerant-based consensus methods might be suitable for the requirement.